\pdfoutput=1
\documentclass[aps,pra,10pt,twocolumn,showpacs,superscriptaddress]{revtex4-1}
\usepackage{amsmath}
\usepackage{latexsym}
\usepackage{amssymb}
\usepackage{bm}
\usepackage{graphicx}
\usepackage{color}

\newcommand{\fishQ}{\mathbf{F}_q}
\newcommand{\fishC}{\mathbf{F}}

\newcommand{\bra}[1]{\langle #1 \vert}
\newcommand{\ket}[1]{\vert #1 \rangle}

\newcommand{\braket}[1]{\langle #1 \rangle}

\newcommand{\tr}{\mbox{tr}}

\begin{document}

\title{Fragile states are better for quantum metrology}
\date{\today}

\author{Kavan Modi}
\email{kavan.modi@monash.edu}
\affiliation{School of Physics and Astronomy, Monash University, Victoria 3800, Australia}

\author{Lucas C. C\'{e}leri}
\affiliation{Instituto de F\'{i}sica, Universidade Federal de Goi\'{a}s, Caixa Postal 131 74001-970, Goi\^{a}nia, Brazil}

\author{Jayne Thompson}
\affiliation{Centre for Quantum Technologies, National University of Singapore, 3 Science Drive 2, Singapore 117543}

\author{Mile Gu}
\affiliation{School of Physical and Mathematical Sciences, Nanyang Technological University, Singapore and Centre for Quantum Technologies, National University of Singapore, Singapore}
\affiliation{Complexity Institute, Nanyang Technological University, Singapore and Centre for Quantum Technologies, National University of Singapore, Singapore}
\affiliation{Centre for Quantum Technologies, National University of Singapore, 3 Science Drive 2, Singapore 117543}

\begin{abstract}
In quantum metrology, quantum probe states are capable of estimating unknown physical parameters to precisions beyond classical limits. What qualities do such states possess? Here we relate the performance of a probe state at estimating a parameter $\phi$ -- as quantified by the quantum Fisher information -- to the amount of purity it loses when $\phi$ undergoes statistical fluctuation. This indicates that the better a state is for estimating $\phi$, the more it decoheres when $\phi$ is subject to noise.
\end{abstract}

\maketitle

\section{Introduction}
Quantum metrology is a promising new quantum technology. It enables techniques for measuring unknown physical parameters at the limits imposed by quantum mechanics \cite{Giovannetti04}. The measurement precision for physical parameters, estimated using traditional interferometric techniques based on single-particle probe states, is subject to the standard quantum limit (also called shot-noise limit). However, recent advances in quantum information science have shown that it is possible to achieve precision beyond this limit by using non-classical probe states \cite{GLM11}. For this reason, a vast amount of effort has been expended into developing a quantum theory for parameter estimation.

Consider some physical process that can encode an unknown parameter $\phi$ onto any incoming state $\rho$, yielding $\rho_\phi$. The goal of quantum metrology is to engineer $\rho$ that are particularly advantageous for determining $\phi$ to high precision. Formally, this is done by measuring the final state $\rho_\phi$ using a positive-operator-valued measure (POVM) $\{\Pi_k\}$ to yield classical probabilities $p_{\phi,k} = \tr[\Pi_k \rho_\phi]$. The precision with which we can estimate $\phi$ is then bounded by the Fisher information
\begin{gather}\label{eq:FisherC}
\Delta \phi \ge \frac{1}{\sqrt{\fishC}}
\quad \mbox{with} \quad
\fishC = \sum_k \frac{[\partial_\phi p_{\phi,k}]^2}{p_{\phi,k}}.
\end{gather}
This inequality is known as the Cram\'{e}r-Rao bound~\cite{cramer, rao} and can be asymptotically saturated~\cite{Cover}. Thus, the greater the Fisher information, the greater the precision to which we can estimate $\phi$ using a quantum probe $\rho$. Consequently the Fisher information is commonly adopted to quantify the performance of quantum metrology. Here we consider unitary processes, where $\rho_\phi = U_\phi \rho U_\phi^\dag$, and $U_\phi = \exp \{-i H \phi\}$ for some Hamiltonian $H$.

What qualities then give quantum probes their operational advantage? A survey of common quantum metrology protocols hints that the states that tend to display quantum mechanical advantage also tend to be rather fragile. The standard quantum metrology protocol, for example, involves the use of NOON states, a highly non-local state on $N$ photons that decoheres completely when a single photon is lost. In this letter we develop methods which allow this intuition to be formalised.

Here we consider the robustness of $\rho$ to stochastic fluctuations on $\phi$. More precisely, $\phi$ is not kept stable, but instead fluctuates with each run of the metrology experiment \cite{PhysRevLett.112.210401}. Mathematically, this can be formalised by introducing a random variable $X$ that takes on some value $x$ with probability $p_x$. Such that, on each run of the experiment, $U_{\phi(x)}$ is applied to $\rho$ with probability $p_x$, creating some $\rho_x  = U_{\phi(x)}^{\dagger}\rho U_{\phi(x)}$. To model gaussian fluctuations, for example, $X$ would be normally distributed about $\phi$ with some variance $(\Delta x)^2$. These fluctuations would then generally induce extra noise on $\rho$, transforming it to a more mixed state $\rho_{\rm avg} = \sum p_x \rho_x$ that has reduced purity. The magnitude of such an effect is then characterised by the purity loss
\begin{gather}
\Delta \gamma = \tr(\rho^2) - \tr(\rho_{\rm avg}^2).
\end{gather}
The purity loss then quantifies the degree in which a state $\rho$ will decohere subject to stochastic fluctuations on $\phi$. Meanwhile the ratio $\Delta \gamma / (\Delta x)^2$ then reflects how fragile such a state is to noise.

Here we develop a number of analytical relations between quantum Fisher information and purity loss. Specifically, we show that the purity loss can bound Fisher information both from above and below, and provide a general construction where the two quantities are exactly related. This indicates that when probing a quantum process with some unknown parameter $\phi$, the precision to which $\phi$ can be determined is fundamentally connected to how stochastic fluctuations in $\phi$ would induce impurity -- the more sensitive an input's purity is to such fluctuations, the better it is at measuring $\phi$.

\begin{figure}[t]
\includegraphics[scale=.3]{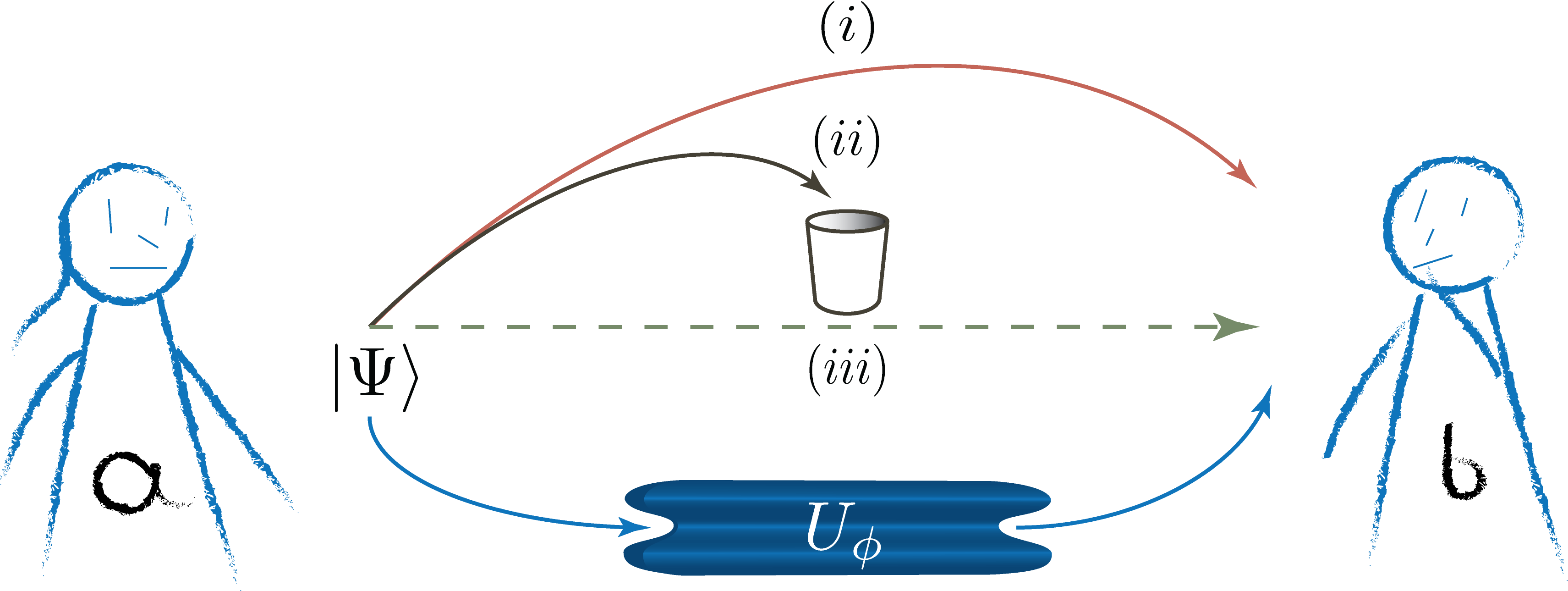}
\caption{Alice prepares a bipartite pure state $\Psi$, and subjects one system to a unitary operation $U = \exp\{-i H \phi\}$. The value of the encoded phase is chosen from a distribution, e.g. see Figure~\ref{Periodic-Gaussian}. We consider three scenarios: (i) Bob receives both the system and the memory; (ii) Bob only receives the system and the memory is fully disposed; (iii) Bob receives the system and Alice makes a measurement on the memory and communicate the outcome to Bob. For each scenario we relate the quantum Fisher information to the linear Holevo quantity, which quantifies the purity loss.
\label{fig:protocol}}
\end{figure}

\section{Framework}
The connection between Fisher information and purity loss can be understood via a family of related protocols between two parties, Alice and Bob. In all scenarios, Alice prepares a (generally) mixed probe state $\rho \in \mathcal{H}_S$ stored in some system $S$. In addition, Alice may also possess a purification of this state on some ancillary memory $M$, such that the joint state of probe and memory
\begin{gather}
\ket{\Psi} = \sum_k \sqrt{q_k} \ket{\alpha_k, k} \quad \mbox{with} \quad \mathrm{Tr}_M (\ket{\Psi}\bra{\Psi}) = \rho.
\end{gather}
Here $\rho = \sum q_k \ket{a_k}\bra{a_k}$.  Alice encodes a phase $\phi(x) = \phi + x$ onto the probe with a unitary transformation $U_{\phi(x)} = \exp\{-i H \phi(x)\}$ on $S$, where $\phi$ is some fixed phase. Meanwhile, we assume that Bob does not know the exact value of $x$, but has pre-knowledge that Alice has encoded a phase that is very close to $\phi$, i.e., $|x| \ll 1$.

Suppose now that in every trial, Alice encodes a fixed $x_0$. Subsequently Alice gives Bob $S$ and possibly $M$ (see variations below), and Bob is challenged to estimate $x_0$. Bob's performance is then gauged by the standard error of his estimate, $\Delta \phi(x_0)$. This framework then exactly captures the standard objective of quantum metrology for estimating an unknown parameter $x_0$.

Specifically, we will consider three variants of the above protocol (See Fig.~\ref{fig:protocol}).
\begin{enumerate}
\item[(i)] Alice does not have $M$, and supplies Bob only $S$.
\item[(ii)]  Alice has $M$, and supplies Bob both $S$ and $M$.
\item[(iii)]  Alice supplies Bob with $S$, but not $M$. Instead she makes a classical measurement on the memory, and communicates the outcome to Bob.
\end{enumerate}
The rationale is that purity loss and the Fisher information will be different depending on how the memory is treated, see \cite{arXiv:1203.0011, micadei}. We will see that just as in uncertainly relations \cite{arXiv:0909.0950}, readout in quantum metrology \cite{micadei}, and quantum illumination \cite{arXiv:1312.3332}, quantum memory plays an important role.

\subsection{Case (i): Using No Quantum Memory}

Case (i) describes the standard quantum metrology protocol. As the memory is discarded, this scenario corresponds to the parameter estimation of the unitary process $U_{\phi(x)}$ using a probe state $\rho$. The standard error of Bob's estimate, $\Delta \phi(x_0)$, is then bounded below by $1/\sqrt{\fishQ (\rho)}$, where
\begin{gather}\label{eq:FisherQ}
\fishQ (\rho)= 2 \sum_{ij} \frac{(\lambda_i - \lambda_j)^2}{\lambda_i + \lambda_j} \left\vert \braket{\psi_i \left| H \right| \psi_j} \right\vert^2.
\end{gather}
is the quantum Fisher information with respect to $\rho$ (Technically the quantum Fisher information is a function of both the probe $\rho$ and the Hamiltonian $H$. For brevity we write $\fishQ(\rho)$ with understanding that the $H$ dependence is implicit). Note that that Quantum Fisher information is obtained by taking the supremum of Eq.~\eqref{eq:FisherC} over all POVMs on $S$, leading to the probabilities $p_{\phi,k}$~\cite{CavesBraunstein94}, and thus is a natural quantum extension of its classical counterpart.

Consider now stochastic fluctuations on $\phi(x)$. Here, the same protocol is repeated over many successive trials, but $x$ is not fixed. In each trial, Alice samples $x$ from some probability density distribution $\mathbf{P}_x$; and applies the corresponding $U_{\phi(x)}$ on the system. This results in a quantum state from the ensemble $\mathcal{E}_\rho := \{\mathbf{P}_x, \, \rho_{x}\}$, where $\rho_{x} := U_{\phi(x)} \, \rho \, U_{\phi(x)}^\dag$. To model stochastic fluctuations, we assume $\mathbf{P}_x$ to be sharply distributed around $\phi$ such that the probability that $|x|$ deviates from zero is negligible (see Figure~\ref{Periodic-Gaussian} of Appendix~\ref{app:PG}), and further, that the specific value of $x$ for each trial is then discarded. Thus Bob returns the ensemble-averaged state
\begin{gather}\label{eq:avgstate}
\rho_{\rm avg} = \int_0^{2\pi} {\rm d}x \, \mathbf{P}_x \, \rho_x.
\end{gather}
The extra noise introduced by these fluctuations can then be characterise by resulting loss of purity
\begin{gather}\label{eq:LH}
\Delta \gamma (\mathcal{E}_\rho)= \tr[\rho^2] - \tr[\rho_{\rm avg}^2],
\end{gather}
where the purity of a state $\rho$ is given by $\tr[\rho^2]$ and the average state $\rho_{\rm avg}$ is given in Eq.~\eqref{eq:avgstate}. This loss of purity then quantifies how fragile $\rho$ is to stochastic fluctuations on $x$.

Equation~\eqref{eq:purrho} in Appendix~\ref{app:int} gives the purity of the state of stochastic perturbation of $x$. Application of Eq.~\eqref{eq:LH} then gives a purity loss of
\begin{gather}
\Delta \gamma  (\mathcal{E}_\rho) \approx
2 (\Delta x)^2 \left( \tr\left[\rho^2 H^2\right] -\tr\left[(\rho H)^2\right] \right),\label{eq:LHrho}
\end{gather}
where $(\Delta x)^2 = \int {\rm d}x \, x^2 \, \mathbf{P}_x - \left(\int {\rm d}x \, x \, \mathbf{P}_x\right)^2$ is the variance of distribution $\mathbf{P}_x$. The approximation in the last equation can be made as tight as desired by making $\mathbf{P}_x$ as sharper and sharper.

We wish to emphasise that $\Delta x$ is very different from $\Delta \phi(x)$ for a fixed $x$. This is because $\Delta x$ represents the strength of the stochastic noise on $x$, and is a quantity that does change with the number of trails.  For a fixed $x$, the variance $\Delta \phi(x)$, on the other hand, represents the stand error in Bob's estimate, progressively becomes smaller as more data are gathered.

It is easy to show that the term on the right is a lower bound to $\fishQ(\rho)$. If we set $1/(\lambda_i + \lambda_j) = 1$ for all $i,\, j$  in Eq.~\eqref{eq:FisherQ},
\begin{gather}\label{eq:lowFish}
\fishQ(\rho) \ge
4 \left(\tr[\rho^2 H^2] - \tr[(\rho H)^2] \right).
\end{gather}
See Appendix~\ref{app:Bounds} for details. This is our first result; we have lower bounded quantum Fisher information in terms of the purity loss
\begin{gather}\label{eq:lowBound}
\fishQ(\rho) \geq 2 \frac{\Delta \gamma (\mathcal{E}_\rho)} {(\Delta x)^2}.
\end{gather}
Here, the ratio $\frac{\Delta \gamma (\mathcal{E}_\rho)} {(\Delta x)^2}$ captures how fragile the purity of $\rho$ with respect to stochastic noise on $x$. Thus, a state -- provided its purity is sufficiently fragile -- is guaranteed to be a more effective quantum probe.

\subsection{Case (ii): Using Quantum Memory}
The second scenario describes the case where the purification is available. That is, Alice gives both $S$ and $M$ to Bob. This allows Bob to improve his estimate of $x_0$ by the use of entangling measurements between $S$ and $M$. Formally, Alice begins with the pure state $\Psi \in \mathcal{H}_S \otimes \mathcal{H}_M$, and encodes a phase with unitary $U_{\phi(x)} \otimes \openone$, generated by the Hamiltonian $H_S \otimes \openone_M$. The Quantum Fisher information for pure states is known to be proportional to the variance of the Hamiltonian~\cite{PhysRevA.87.032324}. Since this Hamiltonian is local to the system, we find
\begin{align}\label{eq:variance}
\fishQ(\Psi) = 4(\tr[\rho H^2] - \tr[\rho H]^2)
=: 4(\Delta H)^2_{\rho},
\end{align}
where $\rho = \tr_{M}[\Psi]$ as before. In general, $\fishQ(\Psi) \geq \fishQ(\rho)$. Mathematically, that is the quantum Fisher information is bounded by the variance. Physically, it reflects the fact that Bob's estimate can only get better with extra information.

We can also evaluate the corresponding purity loss. After stochastic fluctuations, Bob would receive the ensemble-averaged state
\begin{gather}\label{eq:Purecode}
\Psi_{\rm avg} = \int_0^{2\pi} {\rm d}x \, \mathbf{P}_x \Psi_x.
\end{gather}
By evaluate its associated purity (see details in Appendix~\ref{app:intPsi}), we an determine the resulting purity loss
\begin{gather}
\Delta \gamma (\mathcal{E}_\Psi) \approx 2 (\Delta x)^2 (\Delta H)^2_\rho \label{eq:LHR},
\end{gather}
Using one of the bounds derived in Appendix~\ref{app:Bounds}, we obtain our second result: The purity loss and the quantum Fisher information are related by
\begin{gather}\label{eq:upBound}
\fishQ(\rho) \le
2 \frac{\Delta \gamma (\mathcal{E}_\Psi)} {(\Delta x)^2}
\approx \fishQ(\Psi).
\end{gather}
Combining this with our results in case (i), we can also bound $\fishQ(\rho)$ from both above and below:
\begin{equation}
2 \frac{\Delta \gamma (\mathcal{E}_\rho)} {(\Delta x)^2} \leq \fishQ(\rho) \leq 2 \frac{\Delta \gamma (\mathcal{E}_\Psi)} {(\Delta x)^2}.
\end{equation}
These results establish that a state's performance as a quantum probe is intimately related to how quickly it loses purity due to noise on $\phi$. For pure states, the correspondence is an approximate equality. For mixed probe states $\rho$, we can establish both lower and upper bounds. The lower bounded is proportional to the purity loss in $\rho$, where the purity loss of its purification of $\rho$ furnishes an upper bound.

It then makes sense that if we only have limited access to the memory, somewhere in between we must get Fisher information itself. Of course, there are many ways to limit the access to the memory, here we give one example.

\subsection{Case (iii): Saturating the bounds}

To obtain an approximate equality between $\fishQ(\rho)$ and some form of purity loss for mixed probe states, we consider the third scenario. Here Alice measures the memory, and classically communicates the outcome of the measurement to Bob. This measurement generally induces some unravelling of $\rho$ into some ensemble of pure states $\mathcal{Q} =\{q_k, \, \ket{\alpha_k} \}$, such that $\rho= \sum_k q_k \ket{\alpha_k} \bra{\alpha_k}$. This unravelling is of course, not unique, and depends on Alice's choice of measurement.

Regardless, in each run of the experiment, Bob knows exactly which pure state, $\ket{\alpha_k}$, was initially prepared. Thus, for all intents and purposes, this scenario is equivalent to the case where the initial probe state was set to $\ket{\alpha_k}$ with probability $q_k$. We can therefore evaluate the expected purity loss due to stochastic fluctuations individually evaluating the purity loss when each  $\ket{\alpha_k}$ is used as a probe, and taking the statistical average.
\begin{gather}
\gamma(\mathcal{Q}) = \sum_k q_k \gamma(\alpha_k) \approx 2 (\Delta x)^2 \sum_k q_k (\Delta H)^2_{\alpha_k}.
\end{gather}
To relate this to $\fishQ(\rho)$, we make use of the result from Ref. \cite{arXiv:1302.5311}. Here, Yu showed that the quantum Fisher information of a mixed state, for unitary phase encoding, is the convex roof of the variance
\begin{gather}
\fishQ(\rho) = \min_{\{q_k, \, \varphi_k \}} \left(4 \sum_k q_k  (\Delta H)^2_{\varphi_k}\right),
\end{gather}
with $\rho=\sum_k q_k \varphi_k$, where $\varphi_k$ are pure states in $\mathcal{H}_S$. This was previously conjectured by T\'{o}th and Petz in \cite{PhysRevA.87.032324} based on analytical and numerical arguments. Moreover, Yu give a systematic way to find the $\varphi_k$ that saturates this minimisation, the details of which we outline in Appendix~\ref{app:ensemble}. Comparing the above equation to $\gamma(\mathcal{Q})$, we immediately see that
\begin{gather}
\fishQ (\rho) \approx 2 \frac{\chi(\mathcal{Q}_{min})}{(\Delta x)^2},
\end{gather}
where $\mathcal{Q}_{min}$ is the unravelling that minimises purity loss. Thus we have found an (approximate) equality between quantum Fisher information and purity loss. Once again, by taking a highly sharp $\mathbf{P}_x$ we can make the approximation in the last equation as tight as desired. 

\section{Discussion}
Here, we studied our capacity to estimate a parameter $\phi$ that parameterises some unitary channel $U_{\phi(x)}$, as quantified by the Fisher information. We demonstrated that this quantity can be bounded both from above and from below by the rate at which the probe state loses purity as we increase the strength of stochastic perturbations on $\phi$ (as quantified by the variance of the perturbation). These bounds become exact equality when the probe state is pure. For mixed probe states $\rho$, an exact relation between rate of purity loss and Fisher information can also be derived by considering specific unravellings of $\rho$. This formally establishes the common wisdom that the quantum states which are most beneficial for metrology are also the ones that are the most susceptible to noise.

This quantitative relation opens interesting possibilities. The purity is not only relatively easy to compute numerically, but can also be experimentally measured using controlled \textsc{swap} gates without the need for full tomography~\cite{ekert2002direct}. As such, its relation with Fisher information can lead to new experimental methods in evaluating the effectiveness of certain quantum probes. Meanwhile, the Fisher information is recently named an effective quantifier of mascroscopicity -- the degree in which quantum behaviour manifests within a given state at the macroscopic scale~\cite{yadin05}. Our relation then indicates that states that are `more quantum' on a macroscopic scale are also more susceptible to noise.

\begin{acknowledgements}
{\bf Acknowledgements.} We acknowledge the financial support from the Brazilian funding agencies CNPq (Grants No. 401230/2014-7, 445516/2014-3 and 305086/2013-8) and CAPES, the Brazilian National Institute of Science and Technology of Quantum Information (INCT/IQ), the Ministry of Education of Singapore, the National Research Foundation (NRF), NRF-Fellowship (Reference No: NRF-NRFF2016-02) and the John Templeton Foundation (Grant No 54914).
\end{acknowledgements}

\bibliography{Fisher-Holevo}

\newpage
\appendix

\section{Periodic Gaussian}\label{app:PG}

\begin{figure}[t]
\includegraphics[scale=.5]{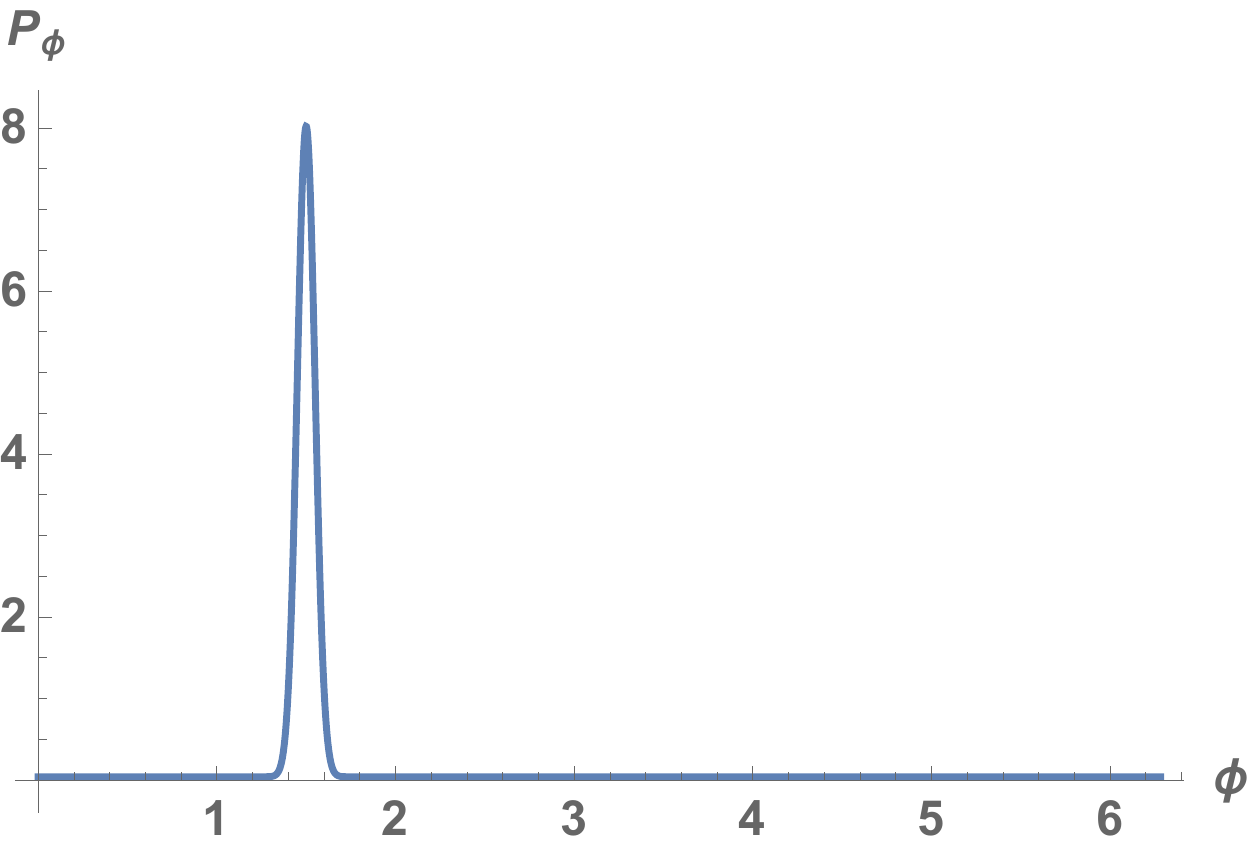}
\caption{A highly peaked Periodic Gaussian distribution. The probability is vanishingly small everywhere except in the vicinity of the peak of $\phi = 1.5$ with $\Delta x = 0.05$. \label{Periodic-Gaussian}}
\end{figure}

Bob's guess for the distribution can be a periodic Gaussian function
\begin{gather}
\mathbf{P}_x	= \frac{1}{\Delta x \sqrt{2 \pi}}
\sum_{k = -\infty}^{ \infty} \exp\left\{-\frac{(x - \phi + 2 \pi k)^2} {2 (\Delta x)^2} \right\}.
\end{gather}
This function is centered at $\phi$ with a variance of $(\Delta x)^2$. We have plotted this function in Figure~\ref{Periodic-Gaussian}.

\section{Computing $\tr[\rho^2_{\rm avg}]$}\label{app:int}

Here we compute the purity of the average state of our protocol (see Eq.~\eqref{eq:LH}).
\begin{align}
\tr[\rho^2_{\rm avg}] =& \tr\left[ \left(\int_0^{2\pi} {\rm d}x \, \mathbf{P}_x \, U_{\phi(x)} \, \rho \, U^\dag_x \right)^2 \right] \\ \notag
=&\tr\left[ \int_0^{2\pi} {\rm d}x \, {\rm d}x' \, \mathbf{P}_x \, \mathbf{P}_x' \, e^{i H x} \, \rho \, e^{-i H (x-x')} \, \rho \, e^{-i H x'} \right]
\end{align}
Plugging in $U_{\phi(x)} = \exp\{-i H x\}$ we get $\tr[\rho^2_{\rm avg}] = \tr\left[\rho \, \mathcal{I} \right]$, where
\begin{gather}
\mathcal{I} = \int_0^{2\pi} {\rm d}x \, {\rm d}x' \, \mathbf{P}_x \mathbf{P}_{x'} e^{i H (x'-x)} \, \rho \, e^{-i H (x'-x)}.
\end{gather}

Here we make use of the fact that $\mathbf{P}_x$ is highly peaked. This means $\mathbf{P}_x \mathbf{P}_{x'} \approx 0$ everywhere except when both $x$ and $x'$ are in the neighbourhood peak of the distribution $\phi$
\begin{gather}
\mathcal{I} \approx \int_{\phi-\epsilon}^{\phi+\epsilon} {\rm d}\phi \, {\rm d}x' \, \mathbf{P}_x \mathbf{P}_{x'} \, e^{i H (x'-x)} \, \rho \, e^{-i H (x'-x)}
\end{gather}
See Figure~\ref{Periodic-Gaussian} for an illustration. Now we Taylor expand the unitary operations as
\begin{gather}
e^{i H (x'-x)} = \sum_{m} \frac{[i H (x-x')]^m}{m!}.
\end{gather}
Since $x-x'$ is small we now discard terms $(x-x')^n$ terms for $n\geq 3$.
\begin{widetext}
\begin{align}
\mathcal{I} \approx&  \int_{\phi-\epsilon}^{\phi+\epsilon} {\rm d}x \, {\rm d}x' \, \mathbf{P}_x \, \mathbf{P}_x' \left( \rho - i [\rho, H] (x'-x)  + H \rho H (x'-x)^2 - \{H^2,\rho\} \frac{(x'-x)^2}{2} \right) + \mathcal{O}((x-x')^3)\\
=& \int_{\phi-\epsilon}^{\phi+\epsilon} {\rm d}x \, {\rm d}x' \, \mathbf{P}_x \, \mathbf{P}_x' \left(\rho - i [\rho, H] (x'-x) + \left[H \rho H - \frac{1}{2} H^2 \rho - \frac{1}{2} \rho H^2\right] ({x'}^2-2 x' x +x^2) \right)\\
=&\rho - i [\rho, H] (\braket{x'}-\braket{x})
+ \left[H \rho H - \frac{1}{2} H^2 \rho - \frac{1}{2} \rho H^2\right] (\braket{{x'}^2}-2 \braket{x'} \braket{x} + \braket{x^2}) ).
\end{align}
\end{widetext}
Since $x$ and $x'$ are dummy variables we have
\begin{gather}\label{eq:I}
\mathcal{I} \approx \rho + 2 \left[H \rho H - \frac{1}{2} H^2 \rho - \frac{1}{2} \rho H^2\right] (\Delta x)^2.
\end{gather}
And, consequently
\begin{align}\label{eq:purrho}
\tr[\rho_{\rm avg}^2] \approx& \tr[\rho^2] - 2 (\Delta x)^2 \left( \tr\left[\rho^2 H^2 \right] -\tr\left[(\rho H)^2\right] \right).
\end{align}

\subsection{Purity of $\Psi_{\rm avg}$}\label{app:intPsi}

Now we perform the same calculations for the purification of $\Psi$. Here we obtain $\tr[\Psi_{\rm avg}^2] = \tr[\Psi \mathcal{J}]$, where
\begin{gather}\notag
\mathcal{J} = \int_0^{2\pi} {\rm d}x \, {\rm d}x' \, \mathbf{P}_x \mathbf{P}_{x'} e^{i H \otimes \openone (x'-x)} \, \Psi \, e^{-i H \otimes \openone (x'-x)}.
\end{gather}
We can simply replace $\rho \to \Psi$ and $H \to H_{SM} = H \otimes \openone$ in Eq.~\eqref{eq:I} to get
\begin{gather}
\mathcal{J} \approx \Psi + 2 \left[H_{SM} \Psi H_{SM} - \frac{1}{2} H_{SM}^2 \Psi - \frac{1}{2} \Psi H_{SM}^2\right] (\Delta x)^2.\notag
\end{gather}
And, consequently
\begin{widetext}
\begin{gather}
\tr[\Psi_{\rm avg}^2] \approx \tr[\Psi^2] - 2 (\Delta x)^2 \left( \tr\left[\Psi^2 H_{SM}^2 \right] -\tr\left[(\Psi H_{SM})^2\right] \right)
= \tr[\Psi^2] - 2 (\Delta x)^2 \left( \braket{\psi |  H^2 \otimes \openone| \psi}  -\braket{\psi | H \otimes \openone | \psi}^2 \right).\notag
\end{gather}
\end{widetext}
Next, we take the partial trace with respect to $M$
\begin{align}
\tr[\Psi_{\rm avg}^2] \approx& \tr[\Psi^2] - 2 (\Delta x)^2 \left( \tr \left[\rho \, H^2 \right] -\tr \left[ \rho \, H \right]^2 \right) \notag\\
 =& \tr[\Psi^2] - 2 (\Delta H)^2_\rho. {(\Delta x)^2}\label{eq:purPsi}
\end{align}

\section{Bounds on $\fishQ(\rho)$}\label{app:Bounds}

An upper bound on $\fishQ(\rho)$ can be obtained by writing $(\lambda_i - \lambda_j)^2 = (\lambda_i + \lambda_j)^2 -4\lambda_i \lambda_j$ in Eq.~\eqref{eq:FisherQ}, leading to
\begin{align}
\fishQ(\rho) =& 2 \sum_{ij} \left(\lambda_i + \lambda_j \notag -\frac{4\lambda_i \lambda_j}{\lambda_i + \lambda_j} \right) \left|\braket{\psi_i \left| H \right| \psi_j} \right|^2 \\
\le& 4 \sum_{i} \lambda_i \braket{\psi_i \left| H^2 \right| \psi_i} -8 \sum_{ij} \lambda_i \lambda_j \left|\braket{\psi_i \left| H \right| \psi_j} \right|^2 \notag\\
=& 4 \tr[\rho H^2]-8 \,{\tr\left[(\rho H)^2\right]}
\le 
4(\Delta H)^2_\rho = \fishQ(\Psi).
\end{align}
Above the first inequality is obtained by noting that $1/(\lambda_i + \lambda_j) \ge 1$. By setting $1/(\lambda_i + \lambda_j) = 1$ we get a lower bound
\begin{align}
\fishQ(\rho) \ge& 2 \sum_{ij} (\lambda_i - \lambda_j)^2 \left|\braket{\psi_i \left| H \right| \psi_j} \right|^2 \\
=& 4 \sum_{ij} \lambda^2_i \braket{\psi_j \left| H \right| \psi_i} \braket{\psi_i \left| H \right| \psi_j}\\
&- 4 \sum_{ij} \lambda_i \lambda_j  \braket{\psi_i \left| H \right| \psi_j} \braket{\psi_j \left| H \right| \psi_i} \notag\\
=
&4 \left(\tr[\rho^2 H^2] - \tr[(\rho H)^2] \right).
\end{align}

\section{Yu's ensemble}\label{app:ensemble}
In \cite{arXiv:1302.5311}, Yu constructs an ensemble $\mathcal{Q} = \{q_k, \ket{\alpha_k} \}$, such that $\rho = \sum_k q_k \ket{\alpha_k}\bra{\alpha_k}$ and, the average variance of $\mathcal{Q}$ is the quantum Fisher information~\footnote{Here we take a slightly different approach than Yu, but we are still following his prescription.}. The average variance of an ensemble is defined as the variance of each pure state $\ket{Q_k}$ averaged over the distribution $q_k$:
\begin{align}
(\Delta H)^2_{\mathcal{Q}} =& \sum_k q_k (\Delta H)^2_{\alpha_k} \notag \\
=& \sum_k q_k \braket{\alpha_k \vert H^2 \vert \alpha_k} - \sum_k q_k \braket{\alpha_k \vert H \vert \alpha_k}^2 \notag\\
=&  \, \tr[\rho H^2] - \sum_k q_k \braket{\alpha_k \vert H \vert \alpha_k}^2. \label{EqAvgVar}
\end{align}

On the other hand, the quantum Fisher information is a function of eigenvalues of the Hamiltonian $H$, and of the eigenvalues ($\lambda_i$) and eigenvectors ($\ket{\psi_i}$) of $\rho$. Yu noticed that, by expanding the first term of the quantum Fisher information in Eq.~\eqref{eq:FisherQ}, this quantity be written as
\begin{align}
\fishQ =& 2 \sum_{ij} (\lambda_i + \lambda_j) \left|\braket{\psi_i \left| H \right| \psi_j} \right|^2 \notag \\
&-2 \sum_{ij} \frac{4\lambda_i  \lambda_j}{\lambda_i + \lambda_j} \left|\braket{\psi_i \left| H \right| \psi_j} \right|^2 \notag \\
=& 4 \, \tr[\rho H^2] - 4 \, \tr[Z^2], \label{EqZ}
\end{align}
where Yu defined
\begin{gather}
Z=\sum_{ij} \sqrt{\frac{2 \lambda_i \lambda_j}{\lambda_i+\lambda_j}} \ket{\psi_i}\braket{\psi_i \vert H \vert \psi_j}\bra{\psi_j}.
\end{gather}

Now, it is clear that if the last terms for Eqs.~\eqref{EqAvgVar} and~\eqref{EqZ} are the same, i.e.,
\begin{gather}\label{EqToProve}
\sum_k q_k \braket{\alpha_k \vert H \vert \alpha_k}^2 = \tr[Z^2]
\end{gather}
then the average variance of the ensemble $\mathcal{Q}$ is the quantum Fisher information: $4 (\Delta H)^2_{\mathcal{Q}} = \fishQ$.

All ensembles that generate the density operator $\rho = \sum_i \lambda_i \ket{\psi_i} \bra{\psi_i} = \sum_k q_k \ket{\alpha_k}\bra{\alpha_k}$ are unitarily connected due to the Jaynes-Hughston-Jozsa-Wootters theorem~
\cite{Schrodinger, PhysRev.108.171, Hughston199314}:
\begin{gather}\label{EqHJW}
\ket{\alpha_k}=\frac{1}{\sqrt{q_k}}\sum_i \sqrt{\lambda_i} \alpha_{ki} \ket{\psi_i}
\end{gather}
where the probabilities $q_k$ are $q_k = q_k \braket{\alpha_k \vert \alpha_k}
=\sum_{i} \lambda_i \vert \alpha_{ki}\vert^2$. The elements $\alpha_{ki} = \sqrt{\frac{q_k}{\lambda_i}} \braket{\psi_i \vert \alpha_k}$ belong to a unitary matrix satisfying
\begin{align}
\sum_k \alpha_{ki} \alpha_{kj}^* =& \sum_k \frac{q_k}{\sqrt{\lambda_i\lambda_j}} \braket{\psi_i \vert \alpha_k}\braket{\alpha_k \vert \psi_j} \notag\\
=& \frac{1}{\sqrt{\lambda_i\lambda_j}} \braket{\psi_i \vert \rho \vert \psi_j} \notag\\
=& \frac{\lambda_i}{\sqrt{\lambda_i\lambda_j}} \braket{\psi_i \vert \psi_j} = \delta_{ij}. \label{EqUni}
\end{align}

In order to construct the ensemble that satisfies Eq.~\eqref{EqToProve} we need to find the elements $\alpha_{ki}$. If we let
\begin{gather}
\label{cond1} Z = \sum_k g_k \Lambda_k ,
\end{gather}
with $g_k = \sqrt{q_k} \braket{\alpha_k \vert H \vert \alpha_k}$ and
\begin{gather}
\label{cond3} \tr[Z \Lambda_k]= \sqrt{q_k} \braket{\alpha_k \vert H \vert \alpha_k},
\end{gather}
then the implication is $\tr[Z^2] = \sum_{k} g_k \tr[Z \Lambda_k] = \sum_k q_k \braket{\alpha_k \vert H \vert \alpha_k}^2$, i.e., Eq.~\eqref{EqToProve} is satisfied. Eq.~\eqref{cond1} will give us $\Lambda_k$, while Eq.~\eqref{cond3} will give us $\alpha_{ki}$.

Let us find $\Lambda_k$ from the condition in Eq.~\eqref{cond1} using Eq.~\eqref{EqHJW}:
\begin{align}
\tr[Z \Lambda_k] =& \sqrt{q_k} \braket{\alpha_k \vert H \vert \alpha_k} \notag\\
=& \sum_{ij} \sqrt{\lambda_i \lambda_j}  \braket{\psi_i \vert H  \vert \psi_j} \frac{\alpha_{ki} \alpha_{kj}^*}{\sqrt{q_k}} \notag\\
=& \sum_{ij} \sqrt{\frac{2\lambda_i \lambda_j} {\lambda_i+\lambda_j}} \braket{\psi_i \vert H  \vert \psi_j}
\frac{\alpha_{ki} \alpha_{kj}^*}{\sqrt{q_k}} \sqrt{\frac{\lambda_i+\lambda_j}{2}}, \notag
\end{align}
where
\begin{align}
\Lambda_k = \sum_{ij}  \frac{\alpha_{ki} \alpha_{kj}^*}{\sqrt{q_k}}  \sqrt{\frac{\lambda_i +\lambda_j} {2}}  \ket{\psi_j}\bra{\psi_i}.
\end{align}


Next, the relationship between $Z$ and $\Lambda_k$ in Eq.~\eqref{cond1} gives us the elements $\alpha_{ki}$:
\begin{align}
Z =& \sum_k g_k \Lambda_k \notag\\
 =& \sum_k g_k \sum_{ij} \frac{\alpha_{kj} \alpha_{ki}^*}{\sqrt{q_k}} \sqrt{\frac{\lambda_i +\lambda_j} {2}}  \ket{\psi_i}\bra{\psi_j} \notag\\
=&  \sum_{ij}  \sqrt{\frac{\lambda_i +\lambda_j} {2}}  \ket{\psi_i}\bra{\psi_j}
\left(\sum_k g_k\frac{\alpha_{kj} \alpha_{ki}^*}{\sqrt{q_k}} \right) \notag\\
=&  \sum_{ij}  \sqrt{\frac{\lambda_i +\lambda_j} {2}}  \ket{\psi_i}\bra{\psi_j} \left(\frac{2 \sqrt{\lambda_i \lambda_j}}{\lambda_i + \lambda_j}
\braket{\psi_i \vert H \vert \psi_j} \right) \notag\\
=&  \sum_{ij}  \sqrt{\frac{\lambda_i +\lambda_j} {2}}  \ket{\psi_i}\bra{\psi_j} \left(\braket{\psi_i \vert Y \vert \psi_j}\right).
\end{align}

For Eq.~\eqref{cond1} to hold, the elements in the parenthesises in line 3 and 4 of the last equation must be the same, coming from the following operator:
\begin{align}
Y =& \sum_{ij} \frac{2 \sqrt{\lambda_i \lambda_j}}{\lambda_i + \lambda_j} \ket{\psi_i} \braket{\psi_i \vert H \vert \psi_j} \bra{\psi_j} \label{EqG}\\
=& \sum_k \frac{g_k}{\sqrt{q_k}} \alpha_{kj} \ket{\psi_j} \bra{\psi_i}\alpha_{ki}^* .
\end{align}
Let $\ket{y_k} = \sum_i \alpha_{ki} \ket{\psi_i}$ and using Eq.~\eqref{EqUni} note that $\braket{y_k \vert y_l} = \sum_{ij} \alpha_{ki} \alpha^*_{lj} \braket{\psi_i \vert \psi_j} = \sum_{i} \alpha_{ki} \alpha^*_{li} = \delta_{kl}$. The implication is that $\ket{y_k}$ are the eigenvectors of $Y$, with eigenvalues $\frac{g_k}{\sqrt{q_k}}$. Moreover, matrix $Y$ in Eq.~\eqref{EqG} can be constructed using $\rho$ and $H$. It can then be diagonalised to obtain $\ket{y_k}$, which gives us $\alpha_{ki} = \braket{\psi_i \vert y_k}$, which is sufficient to construct $\ket{\alpha_k}$ using Eq.~\eqref{EqHJW}.

\end{document}